\documentclass[%
superscriptaddress,
 prl,
rsi,%
amsmath,amssymb,
reprint,%
]{revtex4-1}

\usepackage{graphicx}
\usepackage{dcolumn}
\usepackage{xcolor}
\usepackage{float}

\begin{document}
\author{J. Vester}
\email{jvest@kemi.dtu.dk}
\affiliation{ 
Theoretische Chemie, PCI, Universit{\"a}t Heidelberg, Im Neuenheimer Feld 229, D-69120 Heidelberg, Germany
}%
\affiliation{ 
DTU Chemistry, Technical University of Denmark, 2800 Kgs. Lyngby, Denmark
}
\author{V. Despr\'e}%
\affiliation{ 
Theoretische Chemie, PCI, Universit{\"a}t Heidelberg, Im Neuenheimer Feld 229, D-69120 Heidelberg, Germany
}%
\author{A. I. Kuleff}
\email{alexander.kuleff@pci.uni-heidelberg.de}
\affiliation{ 
Theoretische Chemie, PCI, Universit{\"a}t Heidelberg, Im Neuenheimer Feld 229, D-69120 Heidelberg, Germany
}%

\title{The role of symmetric vibrational modes in the dehoherence of correlation-driven charge migration}

\date{\today}

\begin{abstract}
Due to the electron correlation, a fast removal of an electron from a molecule may create a coherent superposition of cationic states and in this way initiate pure electronic dynamics in which the hole-charge left by ionization migrates throughout the system on an ultrashort time scale. The coupling to the nuclear motion introduces a decoherence that eventually traps the charge and a crucial question in the field of attochemistry is how long the electronic coherence lasts and which nuclear degrees of freedom are mostly responsible for the decoherence. Here, we report full-dimensional quantum calculations of the concerted electron-nuclear dynamics following outer-valence ionization of propynamide, which reveal that the pure electronic coherences last only 2-3~fs before being destroyed by the nuclear motion. Our analysis shows that the normal modes that are mostly responsible for the fast electronic decoherence are the symmetric in-plane modes. All other modes have little or no effect on the charge migration. This information can be useful to guide the development of reduced dimensionality models for larger systems or the search of molecules with long coherence times.
\end{abstract}

\maketitle

\section{\label{sec:level1}Introduction}


With the emergence of the high-harmonic generation techniques, attosecond pulses in the XUV and soft X-ray regime became available, allowing the observation of ultrafast processes occurring on the femto- to attosecond timescale \cite{paul2001observation,corkum2007attosecond}. A natural consequence of the rise of these new experimental capabilities is the development of experimental and theoretical concepts allowing to control and manipulate such processes. One of these new research areas is the attochemistry, aiming at steering the chemical reactivity of a molecular system by controlling its pure electron dynamics \cite{lepine2014attosecond,nisoli2017attosecond,kuleff2018ultrafast}.

A mechanism that lies at the heart of attochemistry is the charge migration \cite{cederbaum1999ultrafast,kuleff2014ultrafast}. The charge-migration phenomenon represents pure electron dynamics being a consequence of the coherent superposition of electronic states, and appears as a charge displacement within an ionized \cite{cederbaum1999ultrafast,breidbach2003migration,kuleff2005multielectron} or excited molecule \cite{dutoi2010Tracing,dutoi2011anExcited}. Contrary to the widely studied charge-transfer process, driven by the nuclear rearrangement, the charge migration does not need any nuclear motion. The process is ultrafast, taking typically just a few femtoseconds, as recently observed for the adenine molecule \cite{maansson2021real,despre2021correlation}, and thus can take place before nuclei have time to move substantially. As the charge distribution in the molecule has a strong impact on the structure and the chemical reactivity of the system, by controlling the charge migration and, for example, localizing the charge on different sites of the molecule, one can steer the system into a desired reaction path. In this context, following the charge-directed reactivity observed in peptide chains \cite{remacle1998charge}, a proof-of-concept example of attochemistry might be the charge migration triggered by ionization. The control of the created positive charge offers the possibility to exert force on the molecular backbone of the system and initiate isomerization, or eventually break or create bonds \cite{valentini2020selective}.

Being a result of initially created electronic coherence, the charge-migration dynamics will be gradually destroyed by the decoherence introduced by the slower nuclear motion. Therefore, an important prerequisite for the realization of attochemistry is the existence of long-lasting electronic coherences giving enough time for achieving control over the charge migration. This question was first discussed for the benzene molecule for which a long-lasting coherence has been predicted \cite{despre2015attosecond}. It has then been shown on several molecules that an ultrafast loss of coherence, in the order of 1-2~fs, can occur \cite{vacher2015electron,jenkins2016nuclear,jenkins2016charge,vacher2017electron,arnold2017electronic}. This raised doubts on the possibility to put attochemistry into practice. However, with the prediction of long-lasting coherences for more molecules \cite{despre2018charge,golubev2020fly,scheidegger2022search} and with the very recent first experimental observation of a long-lasting electronic coherence and its revival for the molecule of silane \cite{matselyukh2021decoherence}, attochemistry appears nowadays as a promising research direction in the field of ultrafast molecular science.  We should also note that very recently a laser-control scheme for manipulating and prolonging the electronic-coherence time was proposed \cite{day2022quantum}.

In order to be able to identify systems of interest for attochemistry, it is necessary to answer the following questions: What favors long-lasting electronic coherences and how can they be predicted for a given molecule? Understanding the molecular properties that favor long-lasting coherence will allow the development of molecular design for ultrafast charge migration \cite{despre2019size,folorunso2021molecular,mauger2022charge,chordiya2022photo}. Knowing which nuclear degrees of freedom are mostly responsible for the loss of electronic coherence will also allow the development of reliable reduced-dimensionality models and thus make possible to study large systems. 

With the help of full-dimensional calculations of the coupled electron-nuclear quantum dynamics initiated by outer valence ionization of the molecule propynamide, here we analyze the vibrational motion causing the coherence loss, that is, the dephasing of the charge-migration oscillations. We show that the symmetric vibrational modes are responsible for the ultrafast loss of coherence and thus those cannot be neglected when constructing reduced-dimensionality models. It appears that a similar behavior of the PES along the symmetric modes is primordial for the existence of long lasting coherences.

\section{\label{sec:meth}Methodology}

As we want to study the coupled electron-nuclear dynamics following ionization of the molecule, we need to go beyond the Born-Oppenheimer approximation. A convenient way to do that is to expand the molecular wave function $\Psi(\mathbf{r},\mathbf{R},t)$ using the full set of the electronic (cationic) eigenstates $\phi_i(\mathbf{r},\mathbf{R})$, or the so-called Born-Huang expansion \cite{born1985international},
\begin{equation}
\Psi(\mathbf{r},\mathbf{R},t) = \sum_i \chi_i(\mathbf{R},t)\phi_i(\mathbf{r},\mathbf{R}),
\label{eq:Born-Huang}
\end{equation}
where $\mathbf{r}$ and $\mathbf{R}$ denote all electronic and nuclear coordinates, respectively. Within this representation, the time-dependent Schrödinger equation transforms into a set of coupled equations for the nuclear wave packets $\chi_i(\mathbf{R},t)$ propagating on the potential-energy surfaces (PES) of the adiabatic electronic states $\phi_i(\mathbf{r},\mathbf{R})$. The derivative couplings between the adiabatic PESs, a consequence of the action of the nuclear kinetic energy operator on the electronic states, are rather expensive computationally, especially when the system has many nuclear degrees of freedom. It is, therefore, convenient to work in the so-called \textit{diabatic} representation, in which the derivative couplings between the electronic states are absent, or at least negligible, but the electronic potentials are directly coupled in the coordinate space. Formally, this means that in the diabatic picture the coupling between the electronic states is moved from the nuclear kinetic energy matrix (which is diagonal in this representation) to the potential energy matrix, which becomes non-diagonal. 

In the cases when no large-amplitude motion of the nuclei is involved, a very powerful method to construct such a diabatic representation is the vibronic-coupling Hamiltonian model \cite{koppel1984multimode}. In this approach, a Taylor expansion around a particular geometry point (usually the Franck-Condon point) is performed in terms of dimensionless (mass- and frequency-weighted) normal-mode coordinates, $q_i$, within a basis of quasi-diabatic states. The Scr\"odinger equation then becomes

\begin{equation}
[(\hat{T}_n + \mathcal{V}_0) \bm{1} + \bm{W}]\chi(\bm{q},t) = i\hbar \frac{\partial\chi(\bm{q},t)}{\partial t},
\end{equation}
where $\hat{T}_n\bm{1}$ and $\mathcal{V}_0\bm{1}$ are the diagonal kinetic and ground-state potential-energy matrices, respectively, while the matrix $\mathbf{W}$ contains the \textit{diabatic} states and the couplings between them.

The kinetic energy operator has the following appearance
\begin{equation}
\hat{T}_n = - \frac{1}{2} \sum_i  \omega_i\frac{\partial^2}{\partial q_i^2}.
\end{equation}
In the case of propynamide, we found that the usual harmonic approximation is not enough and quartic terms have to be included in the Taylor expansion of the zeroth-order diabatic potentials. That is
\begin{equation}
\mathcal{V}_0 = \sum_i [ \frac{1}{2}\omega_iq_i^2 + k_{1i}q_i + \frac{1}{2}k_{2i}q_i^2 + \frac{1}{24}k_{3i}q_i^4].
\end{equation}
In the above expressions, $\omega_i$ is the frequency of normal mode $i$, and $k_{ni}$ are the linear, quadratic, and quartic correction parameters, respectively. The diabatic potential energy matrix $\bm{W}$ contains quadratic diabatic states and linear couplings between them \cite{worth2004beyond}
\begin{equation}
\bm{W_{jj}} = E_j + \sum_i\kappa_i^{(j)}q_i + \sum_i\frac{1}{2}\gamma_i^{(j)}q_i^2
\label{eq:diag}
\end{equation}
\begin{equation}
\bm{W_{jk}} = \sum_i\lambda_i^{(jk)}q_i .
\label{eq:nodiag}
\end{equation}
The energy $E_j$ is the vertical ionization energy of state $j$, while $\kappa_i^{(j)}$ and $\gamma_i^{(j)}$ are the linear and quadratic coupling parameters of mode $i$ in the state $j$, respectively, and $\lambda_i^{(jk)}$ is the linear coupling parameter between states $j$ and $k$ through the coupling mode $i$.

Pure electron dynamics exists as long as there is a coherent superposition of at least two electronic states. Therefore, how long the pure electron dynamics last will depend on the coherence time between the electronic states involved in the coherent superposition. The main source of decoherence between the electronic states is the the nuclear motion. Within the presented above picture of nuclear wave packets propagating on non-adibatically coupled electronic potentials, the decoherence is induced by the decrease of the overlap between the wave packets. This can be caused by the different motion of the wave packets on the different potentials (e.g., they move in different directions) \cite{arnold2017electronic}, as well as by non-adiabatic transfer of population. In cases of large differences in the slopes of the coherently populated potentials, even the initial spread of the wave packet can cause decoherence, as it effectively means a different charge oscillation period for every nuclear geometry within the spread \cite{despre2015attosecond,vacher2015electron}. These effects can be captured by the time-dependent overlap of the wave packets evolving on every pair of electronic states (for more details see \cite{despre2018charge})
\begin{equation}
\chi_{ij}(t) = \langle\chi_i(\bm{q},t)|\chi_j(\bm{q},t)\rangle_{\bm{q}} ,
\label{eq:overlap}
\end{equation}
where $\chi_i(\bm{q},t)$ and $\chi_j(\bm{q},t)$ represent the nuclear wave-packets evolving on the electronic states $i$ and $j$ respectively, and $\langle .|.\rangle_{\bm{q}}$ denotes integration over all nuclear degrees of freedom. 

Another source of decoherence is the interaction of the ionized system with the outgoing electron that can remain close to the system and perturb it \cite{pabst2011decoherence}. This effect is therefore reduced by ionization emitting electrons with high kinetic energy \cite{ruberti2019onset}. The decoherence time certainly depends also on the degree of initially created coherence, or whether the corresponding electronic states are populated simultaneously or with some time-delay. Using ultrashort pulses \cite{chini2014generation} limits the ionization time to just a few attoseconds \cite{maquet2014attosecond} reducing strongly this source of decoherence. In this paper, the sudden-ionization limit will be used, neglecting the decoherence induced by the ionized electron and the ionization time. The decoherence will, therefore, be quantified only through the wave-packet overlap $\chi_{ij}(t)$ given in Eq.~\ref{eq:overlap}.

\section{Results and Discussions}

Propynamide was chosen based on studies of the propiolic acid \cite{despre2018charge} which revealed long-lasting electronic-coherence times of more than 10 fs. Propynamide possesses a similar to propiolic acid structure, in which the carboxylic oxygen is replaced by an amine group. The ionization spectrum of propynamide, and respectively the PESs of the states of interest, has been computed with the third-order non-Dyson algebraic diagrammatic constructions scheme (nD-ADC(3)) \cite{schirmer1998non}. nD-ADC(3) is a Green’s function based method which has the advantage to provide all the cationic eigenstates of the system in a single calculation, whilst other wave-function based methods need to calculate each state separately.

\begin{figure}
\includegraphics[scale=2.0]{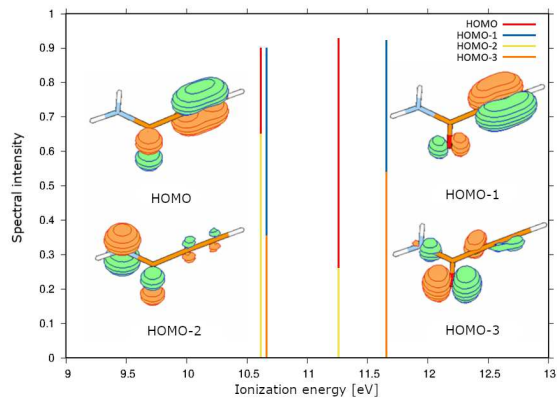}
\caption{\label{fig:spectrum} The outer-valence part of the ionization spectrum of propynamide in its equilibrium geometry computed at nD-ADC(3)/Def2TZVP level of theory. The molecular orbitals involved in the hole mixing in the cationic states are also shown.}
\end{figure}

The outer-valence part of the ionization spectrum of propynamide at the equilibrium geometry of the neutral system is presented in Fig.~\ref{fig:spectrum}. In the spectrum, each line represents a cationic eigenstate of the system and each color gives the contribution the removal of an electron from a given molecular orbital has to this state. The height of the line, or spectral intensity, reflects the mono-electronic part of the state. At third order of perturbation theory, the ADC includes all mono-electronic excitations, 1-hole (1h - removal of an electron from a specific molecular orbital), and all multi-electronic excitation of the type 2-hole-1-particle (2h1p - removal of two electrons from the occupied molecular orbitals and promoting one of them to a virtual orbital). The height of the line, therefore, is given by the sum of the weights of all 1h configurations, while the remaining part up to a value of 1 represents the weight of the 2h1p configurations. 

The energy region presented in the top panel of Fig.~\ref{fig:spectrum} is well described through mono-electronic excitations. However, it appears that the first and the third state, as well as the second and the fourth one are strong mixtures of two pairs of orbitals, i.e. the electron correlation leads to a hole mixing \cite{cederbaum1986correlation,breidbach2003migration,breidbach2007migration}. The ionization of one of the orbitals involved in the hole mixing will lead to a coherent population of the two states involved and will thus trigger a correlation-driven charge migration. We have to note that a very similar hole mixing appears in the first and third state of the propiolic acid (see Refs.~\cite{golubev2017quantum,despre2018charge}). It is thus quite interesting that even after changing the acceptor site from a carboxyl (propiolic acid) to amid (propynamide) group, the strong hole mixing remains. It is also noteworthy that in the propiolic acid only one hole-mixing structure is present, and therefore only one charge migration from the carbon triple bond to the carbonyl oxygen is possible, while in propynamide, two hole-mixing structures appear, offering two possibilities to initiate charge-migration dynamics. The examination of the Hartree-Fock orbitals (also shown in Fig.~\ref{fig:spectrum}) reveals that the first charge migration will also take place mostly between the amine and alkyne groups (HOMO and HOMO$-2$ hole mixing), as in the propiolic acid, while the second one will be mostly between the carbonyl and alkyne groups (HOMO$‑1$ and HOMO$‑3$ hole mixing). With an energy difference between the states involved in the first hole mixing of 0.65~eV, the corresponding dynamics are expected to have a period of 6.4~fs. The energy difference in the second structure is 0.99~eV, which will lead to a charge-oscillation period of 4.2~fs.

As explained in the methodology section, in order to include the nuclear dynamics and examine their influence on the pure electronic charge-migration oscillations, we need to construct a vibronic-coupling Hamiltonian. For this purpose, we first computed the PESs of the 4 states of interest along the 18 vibrational modes of propynamide using the nD-ADC(3) method, and the ground-state PES using the MP2 method. The normal modes of the neutral molecule were obtained at DFT/B3LYP level of theory and can be found in Tab.~\ref{tab:vibrational_modes}. All these calculations have been performed with Def2TZVP basis set \cite{weigend2006accurate,weigend2005balanced}. Such constructed adiabatic potentials were used as a reference for obtaining the parameters of the vibronic-coupling Hamiltonian. As the transformation from a dibatic to an adiabatic electronic-state basis is performed with the eigenvector matrix of the diabatic potential-energy matrix, by diagonalizing the latter we should recover the adiabatic PESs. One, therefore, needs to find the parameters of the vibronic-coupling model, Eqs.~(\ref{eq:diag})-(\ref{eq:nodiag}), such that when the diabatic potential-energy matrix is diagonalized, the adiabatic PESs are reproduced as good as possible.

\begin{table}
\centering
\caption{Vibrationnal modes of propynamide calculated at the B3LYP/Def2TZVP level.}
\label{tab:vibrational_modes}
\begin{tabular}{ccc} 
\hline \hline
Mode & $\nu$~[cm$^{-1}$] & Symmetry   \\
\hline
1    & 187.8                              & a'        \\ 
2    & 210.4                              & a''       \\ 
3    & 250.6                              & a''       \\ 
4    & 491.2                              & a'        \\ 
5    & 564.2                              & a''       \\ 
6    & 600.0                              & a'        \\ 
7    & 689.2                              & a'        \\ 
8    & 718.7                              & a''       \\ 
9    & 765.9                              & a''       \\ 
10   & 797.1                              & a'        \\ 
11   & 1092.6                             & a'        \\ 
12   & 1335.4                             & a'        \\ 
13   & 1615.0                             & a'        \\ 
14   & 1748.9                             & a'        \\ 
15   & 2214.9                             & a'        \\ 
16   & 3466.4                             & a'        \\ 
17   & 3582.0                             & a'        \\ 
18   & 3716.7                             & a'        \\
\hline\hline
\end{tabular}
\end{table}

\begin{figure}
\includegraphics[scale=1.0]{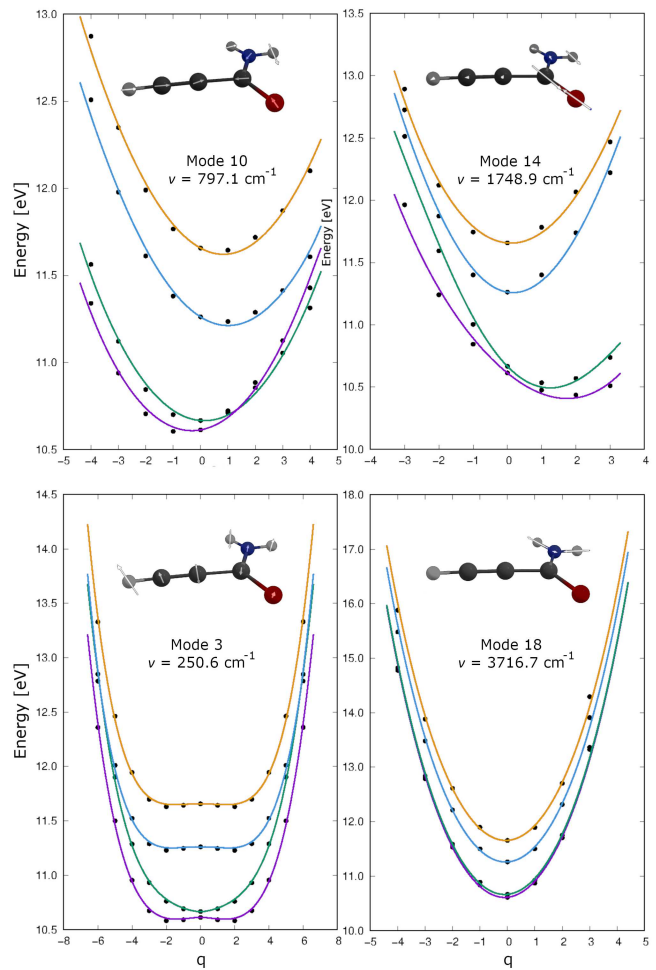}
\caption{\label{fig:PES} Potential-energy surfaces of the first four cationic eigenstates of propynamide along normal modes 3 (250.6~cm$^{-1}$), 10 (797.1~cm$^{-1}$), 14 (1748.9~cm$^{-1}$), and 18 (3716.7~cm$^{-1}$). The adiabatic PESs obtained with nD-ADC(3)/Def2TZVP are depicted with dots, and those obtained by diagonalizing the diabatic potential-energy matrix of our vibronic-coupling Hamiltonian are depicted with solid lines.}
\end{figure}

This was done with the VCHam program \cite{beck2000multiconfiguration} using a least-square fit algorithm. Typical examples of the result of this procedure are shown in Fig.~\ref{fig:PES}, where the adiabatic PESs and their reconstruction from the diabatic potential-energy matrix are depicted for normal modes 3, 10, 14 and 18. We see that our vibronic-coupling Hamiltonian model describes very well the PES of the four states of interest along the four depicted modes (we get the same agreement along the remaining normal modes). We also see from the example of mode 3 why the inclusion of a quartic term in the Hamiltonian was necessary. The modes have been chosen for their importance in the following discussion.

With the vibronic-coupling Hamiltonian we can now propagate an initial state of interest on the four non-adiabatically coupled cationic states along all 18 nuclear degrees of freedom. This was done using the multi-configuration time-dependent Hartree (MCTDH) algorithm \cite{meyer1990multi,beck2000multiconfiguration} with the Heidelberg MCTDH Package \cite{mctdhPackage}. The initial state was constructed assuming a sudden removal of an electron from HOMO or from HOMO$-1$, from a vibrationally cold neutral molecule. The initial state $|\psi(t_0)\rangle$ is thus obtained by projecting the ground-state vibrational wave function on the four cationic states using the HOMO and HOMO$‑1$ hole-mixing parameters as weights along all nuclear degrees of freedom. For each state, the hole-mixing parameters along all 18 normal modes have been extracted from the nD-ADC(3) calculations. After the ionization of an electron from either HOMO or HOMO$-1$, the charge is mainly localized on the alkyne group. Due to the hole mixing between states 1 and 3, as well as between states 2 and 4, it is expected that the charge will migrate to the amine and to the carbonyl group, respectively (see also Fig.~\ref{fig:spectrum}). In case we have a long-lasting electron coherence, the charge will have time to perform several or even many oscillations between these two sites. 

\begin{figure}
\includegraphics[scale=0.34]{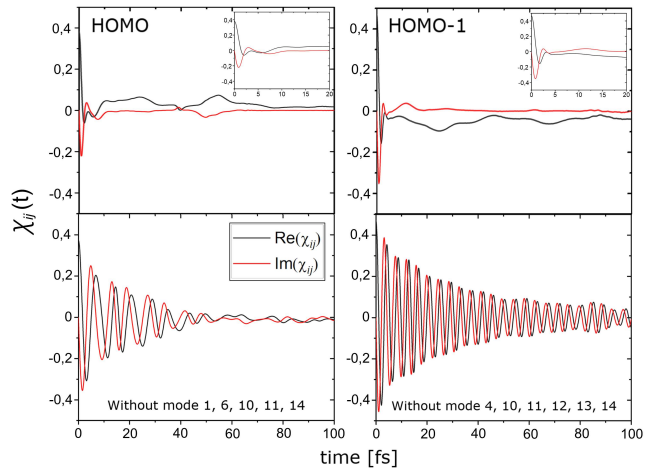}
\caption{\label{fig:overlap} Time-dependent overlap $\chi_{ij}(t)$ between the nuclear wave packets propagating on states $i$ and $j$. The real part of $\chi_{ij}$ is depicted in black and the imaginary one in red.
Top panel: Dynamics initiated by removing an electron from the HOMO (left) and HOMO$-1$ (right) taking all 18 normal modes into account. The insets show the first 20~fs. 
Bottom panel: Dynamics initiated by removing an electron from the HOMO, excluding modes 1, 6, 10, 11, and 14 (left) and dynamics initiated by removing an electron from the HOMO$-1$, excluding modes 4, 10, 11, 12, 13, and 14 (right).}
\end{figure}

As we discussed above, the electron-coherence time can be determined by the time-dependent overlap between the wave packets propagating on the respective pair of states, Eq.~(\ref{eq:overlap}), that is, $\chi_{13}(t)$ if we start by ionizing the HOMO, and $\chi_{24}(t)$ if the initial state is prepared by removing of an electron from HOMO$-1$. The time-evolution of these quantities are shown in the top left and right panel of Fig.~\ref{fig:overlap}, respectively. We see that for both initial states, $\chi_{ij}(t)$ decreases very fast, meaning that the electron coherence is lost within the first 2-3~fs. The quantity $\chi_{ij}(t)$ does increase again between 10‑80~fs, due to partially overlapping portions of the wave packets, but no oscillations appears meaning that the electron dynamics are nearly absent. It appears, therefore, that in contrast to the propiolic acid, the electron coherence in propynamide is lost extremely fast and the charge does not have time to undergo even a single full oscillation, in both considered cases -- ionization out of HOMO and ionization out of HOMO$-1$.

What causes this ultrafast decoherence? If particular modes are responsible for it, understanding their characteristics will permit to better determine which modes are necessary to be taken into account such that we can construct reduced-dimensionality models that correctly capture the non-adiabatic dynamics and the electron-coherence timescale. This will greatly simplify the study of larger systems and will eventually allow to design molecules exhibiting long-lasting coherence. 

We therefore performed reduced-dimensionality calculations of propynamide by excluding different sets of normal modes until long electronic coherences were observed. The results of these simulations are shown in the bottom panels of Fig.~\ref{fig:overlap}. We see that electronic coherences are present for more than 30~fs after ionization of the HOMO, and for more than 80~fs after ionization of the HOMO$-1$ if 5, respectively 6 modes are neglected. The normal modes that induce the strong decoherence in the case of removing an electron from the HOMO are modes 1, 6, 10, 11, and 14. In the case of ionization from HOMO$-1$, the most relevant modes are 4, 10, 11, 12, 13, and 14. The fast decoherence introduced by modes 10 and 14 can be understood by looking at Fig.~\ref{fig:PES}. After the creation of the wave packets at the Franck-Condon region (around $q=0$), the wave packets on the lower two states will evolve in opposite direction from the wave-packets on the two upper states, reducing in that way the respective overlap on an ultrashort time scale.

The different reduced-dimensionality dynamics simulations reveal that there is a specific set of modes that are mostly responsible for the loss of coherence, i.e., modes that lead to a fast decrease of the overlap between the nuclear wave packets evolving on the respective electronic states. The modes of interest that induce decoherence are modes 1, 4, 6, 10, 11, 12, 13, and 14. What all these modes have in common is that they are symmetric (in-plane) modes. This can be rationalized as follows. Contrary to the non-symmetric modes, the symmetric ones have the characteristic of behaving differently with respect to a positive or negative displacement from the equilibrium position ($q=0$). That is, the potentials are not symmetric around $q=0$. This leads to overall steeper slopes of the potentials, often with opposite gradients, and thus to a greater variation of the energy difference between the states even at the Frack-Condon region compared to the non-symmetric modes. This characteristic topology of the states along the symmetric modes is responsible for the ultrafast loss of electron coherence. It is also important to mention that the ultrafast loss of coherence in propynamide is mostly due to the low- and intermediate-frequency symmetric modes. Modes of high frequency (hydrogen vibrations) do not introduce strong decoherence, as they have a smaller impact on the shape of the PES. Therefore, when constructing reduced-dimensionality models that can well capture the short-time coupled electron-nuclear dynamics, one should try to include as many symmetric modes of intermediate and low frequency as possible. In contrast, the non-symmetric modes, as well as the symmetric ones of high frequency, could be neglected.

\section{Conclusion}

In this study, we examined the concerted electron-nuclear dynamics in propynamide initiated by an outer-valence ionization through full-dimensional quantum calculations. The adiabatic PESs were calculated at the nD-ADC(3) level, and the dynamics in the first four cationic states along all 18 normal modes were obtained through the propagation of the respective initial-state nuclear wave packets with a specifically constructed vibronic-coupling Hamiltonian using the MCTDH algorithm. The aim was to find out how long the pure electronic coherences created by ionization out of the HOMO or the HOMO$-1$ last and identify the nature of the vibrational modes mostly responsible for the loss of coherence. The motivation was that this information could guide the development of reduced-dimensionality models for larger systems, for which accounting for all nuclear degrees of freedom is out of reach.

The simulations with all normal modes included show that the electronic coherences last only 2‑3~fs, and the hole charge does not have time to undergo even a single complete oscillation. The additional reduced-dimensionality dynamical simulations reveal that only symmetric (in-plane) modes induce decoherence. Moreover, only the low and intermediate frequency modes lead to an ultrafast loss of coherence. Therefore, it is important to include as many symmetric modes of intermediate and low frequency as possible when constructing reduced-dimensionality models. The non-symmetric ones and the symmetric modes of high frequency could be neglected as they do not contribute to the fast decoherence.

\begin{acknowledgments}
The authors thank Nikolay Golubev for many valuable discussions. VD and AIK acknowledge financial support from the DFG through the QUTIF priority programme.
\end{acknowledgments}

\bibliography{propynamide.bib}

\end{document}